\documentclass[twocolumn, superscriptaddress, 10pt]{revtex4-2}

\usepackage[T1]{fontenc}

\usepackage[colorlinks=true,linkcolor=blue,urlcolor=blue,citecolor=blue,breaklinks=true]{hyperref}

\usepackage{amsmath, amsthm, amssymb, bbm, mathtools}

\usepackage{graphicx}

\renewcommand{\vec}[1]{\boldsymbol{#1}}
\newcommand{\matr}[1]{\boldsymbol{#1}}
\newcommand{\eye}{\mbox{$\mbox{1}\!\mbox{l}\;$}}

\newtheorem{thm}{Theorem}
\newtheorem{defn}[thm]{Definition}
\newtheorem{lemma}{Lemma}
\newtheorem{corr}{Corollary}

\begin{document}

\title{Stability bounds of droop-controlled inverters in power grid networks}

\author{Philipp C. B\"ottcher}
\email[Corresponding author: ]{p.boettcher@fz-juelich.de}
\affiliation{Forschungszentrum J\"ulich, Institute of Energy and Climate Research, Systems Analysis and Technology Evaluation (IEK-STE), 52428 J\"ulich, Germany}
\affiliation{Forschungszentrum J\"ulich, Institute of Energy and Climate Research, Energy Systems Engineering (IEK-10), 52428 J\"ulich, Germany}

\author{{Leonardo Rydin Gorj\~ao}}
\affiliation{Faculty of Science and Technology, Norwegian University of Life Sciences, 1432 Ås, Norway}

\author{Dirk Witthaut}
\affiliation{Forschungszentrum J\"ulich, Institute of Energy and Climate Research, Systems Analysis and Technology Evaluation (IEK-STE), 52428 J\"ulich, Germany}
\affiliation{Forschungszentrum J\"ulich, Institute of Energy and Climate Research, Energy Systems Engineering (IEK-10), 52428 J\"ulich, Germany}
\affiliation{Institute for Theoretical Physics, University of Cologne, 50937 Köln, Germany}

\begin{abstract}
The energy mix of future power systems will include high shares of wind power and solar PV.
These generation facilities are generally connected via power-electronic inverters.
While conventional generation responds dynamically to the state of the electric power system, inverters are power electronic hardware and need to be programmed to react to the state of the system.
Choosing an appropriate control scheme and the corresponding parameters is necessary to guarantee that the system operates safely.
A prominent control scheme for inverters is droop control, which mimics the response of conventional generation.
In this work, we investigate the stability of coupled systems of droop-controlled inverters in arbitrary network topologies.
Employing linear stability analysis, we derive effective local stability criteria that consider both the overall network topology as well as its interplay with the inverters' intrinsic parameters.
First, we explore the stability of an inverter coupled to an infinite grid in an analytic fashion and uncover stability and instability regions.
Secondly, we extend the analysis to a generic topology of inverters and provide mathematical criteria for stability and instability of the system.
Last, we showcase the usefulness of the criteria by examining two model systems using numerical simulations.
The developed criteria show which parameters might lead to an unstable operating state.  
\end{abstract}

\maketitle

\section{Introduction}
The ongoing transition to a sustainable energy system challenges the stability of electric power systems in several ways~\cite{Sims11}. 
Renewable power sources, such as wind power and solar photovoltaics, fluctuate on a large range of time scales~\cite{Anvari16, Schafer18, weber2019wind}.
Wind power is often located at favorable locations far away from the centers of the load, requiring long-distance transmission~\cite{Pesc14}.
Cellular or microgrid concepts are being developed to increase robustness, enabling small or islanded power systems to operate safely~\cite{lopes2006defining, Rocabert2012}.
Finally, wind turbines and solar panels are generally connected to the power grid via power-electronic inverters instead of traditional synchronous machines, which fundamentally alters frequency and voltage dynamics~\cite{teodorescu2011grid, Milano2018}.
A system driven by inverter-connected generation is fundamentally different from present systems around the world which still have a considerable amount of generation by large rotating generator masses.
While conventional generation by large rotating masses is synchronously coupled to the grid due to electromechanical properties, inverters can be freely programmed to react to a measured state of the connected power grid.
Different control algorithms have been proposed to achieve different targets.
Studying how the dynamics of inverters are affected by different endogenous factors, e.g., different line types and model details~\cite{HenriquezAuba2020, HenriquezAuba2022}, becomes increasingly important as we move to a system dominated by power-electronic inverters, particularly by droop-controlled inverters~\cite{SimpsonPorco2013, Schiffer2019, Matveev2020, Bai2022}.
Furthermore, it is essential to understand the collective dynamics of coupled inverters as well as how their interaction with the topology of the underlying power grid influences the set of system parameters that lead to a stable operating state. 
For example, the existence of different network motifs and topologies can destabilize the grid~\cite{Jaros2023}, induce multistability~\cite{Hellmann2020, balestra2019multistability}, or influence the propagation of disturbances~\cite{tamrakar2018propagation}.

This article presents a set of analytically derived stability criteria for droop-controlled inverters.
It extends upon previous studies on the topic, especially the ones by Schiffer~\textit{et al.}, Refs.~\cite{Schiffer2014, Schiffer2016}, and Ref.~\cite{bottcher2022dynamic} dealing with synchronous machines, by providing a discussion of the interplay of frequency, voltage dynamics, the power-grid topology, and their impact on the stability of the full system.
Additionally, we make use of these criteria to derive a set of necessary and sufficient conditions for the linear stability of power grids with an arbitrary topology.
In particular, we derive upper bounds for the reactive power droop gains and lower bounds for the connectivity of the network.

The paper is organized as follows.
In Sec.~\ref{sec:model_inverter_based_micro}, we introduce the model used to describe the collective dynamics of inverter-based networks with the dynamics of droop-controlled inverters being described in Sec.~\ref{sec:model_droop_controlled_inv}, the equations pertaining to the connecting power grid in Sec.~\ref{sec:model_network_eqs}, and the stable states of operation or fixed points in Sec.~\ref{model_fixed_pts}.
In Sec.~\ref{sec:single_inverter}, the dynamics of a single inverter coupled to an infinite grid are investigated to ascertain the existence of different regions of stability for different system parameters.
Subsequently, we extend the stability analysis to arbitrary network topologies by constructing the Jacobian of the full system and formulating conditions for stability and instability in Sec.~\ref{sec:linear_stab_networks}.
In Sec.~\ref{sec:explicit_conditions}, we extend these conditions to derive a set of explicit stability conditions expressed as inequalities.
Finally, in Sec.\ref{sec:numerical_results}, we test the tightness of the derived conditions and thus their usefulness in determining the stability of considered fixed points by numerical experiments in two test systems. 

\section{Modeling inverter-based power grids}
\label{sec:model_inverter_based_micro}
\subsection{Dynamics of droop-controlled inverters}
\label{sec:model_droop_controlled_inv}
Traditional power systems rely on synchronous machines to produce power and supply every consumer in the network with electricity.
Synchronous machines possess an intrinsic relation between their power output and their frequency and phase angle, described by the swing equation~\cite{Machowski2020}.
Power-electronic inverters, on the contrary, do not possess such an intrinsic relation \textit{a priori} but offer some flexibility to design their control mechanics and response.
The vast majority of the commonly used inverters employ a control scheme that mimics the dynamics of conventional synchronous machines.
In most cases, a simple proportional control is applied, whereas in fair argumentation the frequency regulation is designed to be an instantaneous reaction and the voltage control is regulated with a delay.
These types of controllers are denoted droop controllers since they `droop' or decrease their internal characteristics (frequency or voltage) to match a desired state of operation.

The basic state variables of a set of inverters $j=1, \ldots, N$ are the voltage magnitude $E_j \in\mathbb{R}_+$ and the voltage phase angle $\delta_j\in\mathbb{S}$.
The control system adjusts these state values according to measurements of the active and reactive power exchanged with the grid.
Following Schiffer~\textit{et al}~\cite{Schiffer2016}, a general control scheme for a inverter $j$ obeys the equations
\begin{equation}
\begin{aligned}
    \dot{\delta}_j &=u_j,\\
    \tau_j^V \dot{E_j}&= - E_j + v_j,
\end{aligned}\label{eq:control_scheme}
\end{equation}
where $\tau_j^V \in\mathbb{R}_+$ is the recovery time of the voltage dynamics under control.
In a simple proportional, or droop, control scheme, the frequency control $u_j\in \mathbb{R}$ is directly proportional to the active power $P^{\mathrm{el}}_j$ and the voltage control $v_j\in\mathbb{R}$ is proportional to the reactive power $Q^{\mathrm{el}}_j$ (in VAR). 
Hence, the control signals read
\begin{equation}\label{eq:linear_control}
\begin{aligned}
  u_j &=\omega^d - \kappa_j\left(P^{\mathrm{mes}}_j - P^{d}_j\right),\\
  v_j &=E_j^d - \chi_j\left(Q^{\mathrm{mes}}_j - Q^{d}_j\right),
\end{aligned}
\end{equation}
and
\begin{equation}\label{eq:linear_control_power}
\begin{aligned}
\tau_j\dot{P}^{\mathrm{mes}}_j &= - P^{\mathrm{mes}}_j + P_j^{\text{el}},\\
\tau_j\dot{Q}^{\mathrm{mes}}_j &= - Q^{\mathrm{mes}}_j + Q_j^{\text{el}},
\end{aligned}
\end{equation}
where the superscript $\cdot^{\mathrm{mes}}$ indicates the measured values of real and reactive power and the superscript $\cdot^{d}$ stands for \textit{desired}, i.e., referring to the desired frequency or voltage at machine $j$.
Naturally, the desired frequency $\omega^d$ the inverters should attain is unique across the power grid, whereas the desired voltage $E_j^d$ depends on each generator.
A natural choice for the desired frequency is the reference frequency of $50\,$Hz or $60\,$Hz, i.e., the most common mains frequencies, which are within a certain range that balances the efficiency of generators and motors and certain requirements of equipment~\cite{Machowski2020}.
The parameters $\kappa_j$ and $\chi_j$ are the droop gains for the active and reactive power, respectively.

The measurement of the active and reactive power is typically not instantaneous.
This is taken into account by a low-pass filter such that the measured values $P^{\mathrm{mes}}_j$ and $Q^{\mathrm{mes}}_j$ at machine $j$ are given by Eq.~\eqref{eq:linear_control_power} with a low-pass-filter time constant $\tau^{V}_j$~\cite{Coelho2002}, where $P_j^{\mathrm{el}}$ and $Q_j^{\mathrm{el}}$ denote the instantaneous values.

We now summarize the equations of motion describing the inverter and its control system.
The measured values of the power $P^{\mathrm{mes}}_j$ and $Q^{\mathrm{mes}}_j$ can be expressed in terms of the control signals using the relations Eq.~\eqref{eq:linear_control}.
A further simplification can be made as the recovery time of the voltage dynamics is typically much lower than the time constant of the low-pass filter such that we can set $\tau_j^V=0$ in Eq.~\eqref{eq:control_scheme}, which yields $v_j = E_j$.
Hence, we obtain the following set of equations of motion of an inverter $j$~\cite{Schiffer2014, Schiffer2016}
\begin{equation}\label{eq:inverters_pre}
\begin{aligned}
    \dot{\delta}_j&=\omega_j,\\
    \tau_j \dot{\omega}_j &= - \omega_j + \omega^{\mathrm{d}} - \kappa_j\left(P^{\mathrm{el}}_j - P^{\mathrm{d}}_j\right),\\
    \tau_j \dot{E}_j &= - E_j+ E_j^{\mathrm{d}} - \chi_j\left(Q^{\mathrm{el}}_j - Q^{\mathrm{d}}_j\right).
\end{aligned}
\end{equation}
The real and reactive power exchanged with the grid, $P^{\mathrm{el}}_j$ and $Q^{\mathrm{el}}_j$, depend on the state of all elements in the grid.
To close the equations of motion we thus have to specify the network equations for the power grid.

\subsection{The network equations}
\label{sec:model_network_eqs}

The real and reactive power flows in a power grid are described by the classical alternating current (AC) load-flow equations~\cite{Machowski2020}.
We consider a network consisting of inverter (active) and load (passive) nodes, which are modeled by a constant impedance to the ground.
The passive nodes can be eliminated using the Ward or Kron reductions resulting in an effective network consisting of inverter nodes only~\cite{Doerfler2011b}.
In the following, we use this reduced network only.
Without loss of generality, we can assume that the network is connected, as we can analyze different components separately otherwise.

The complex voltage at each inverter node $j=1,\ldots,N$ is written as $V_j = E_j e^{i \delta_j}$ with $E_j \in \mathbb{R^+}$ and $\delta_j \in \mathbb{S}$.
The total current injected in the grid is linear in the voltages according to Ohm's law and can be written as
\begin{equation}
    I_{j} = \sum_{\ell=1}^N Y_{j,\ell} V_\ell.
\end{equation}
Here, we have introduced the \emph{nodal admittance matrix} $\mathbf{Y} \in \mathbb{C}^{N\!\times\!N}$ with the entries
\begin{equation}
  Y_{j,\ell} = G_{j,\ell} + i B_{j,\ell}= \left\{ 
   \begin{array}{lll}
   \displaystyle \hat y_{j,j} + \sum_{k\neq j} y_{j,k} &  \mbox{if} & j = \ell; \\ [2mm]
     - y_{j,\ell} & \mbox{if} & j \neq \ell,
   \end{array} \right.
   \label{eq:admittance}
\end{equation}
where $ y_{j,\ell}$ is the admittance between nodes $j$ and $\ell$ in the effective Kron-reduced network.
Furthermore, $\hat y_{j,j} = \hat{g}_{j,j}+i\hat{b}_{j,j}$, where $\hat{g}_{j,j}$ and $\hat{b}_{j,j}$ denote the shunt conductance and susceptance, respectively.
Without the shunts, the matrices $\matr{G}$ and $\matr{B}$ are graph Laplacian matrices \cite{Newman2018}. 

The apparent power feed-in $S_j^{\text{el}}$ at node $j$ is  given by
\begin{equation}
    S_j^{\text{el}} = V_j I_j^* = \sum_{\ell=1}^N V_j Y^*_{j,\ell} V_\ell^*.
\end{equation}
Decomposing into real and reactive power,  $S_j^{\text{el}} = P_j^{\text{el}}+iQ_j^{\text{el}}$, yields
\begin{subequations}\label{eq:power_and_current-lossy}
\begin{align}
    P^{\text{el}}_j &= \sum_{\ell=1}^N E_{j} E_{\ell} 
       \left[B_{j,\ell} \sin(\delta_{j,\ell} ) 
       + G_{j,\ell} \cos(\delta_{j,\ell}  )\right],\label{eq:active_power-lossy} \\
    Q^{\text{el}}_j &= \sum_{\ell=1}^N E_j E_{\ell} 
      \left[-B_{j,\ell} \cos(\delta_{j,\ell}  ) 
       + G_{j,\ell} \sin(\delta_{j,\ell}  )\right],\label{eq:reactive_power-lossy} 
\end{align}
\end{subequations}
using the shorthand $\delta_{j,\ell} = \delta_{j} - \delta_{\ell}$.
In the main part of the manuscript, we restrict ourselves to lossless grids setting $G_{j,\ell}$ to zero. 
The network equations then read
\begin{subequations}\label{eq:power_and_current}
\begin{align}
    P^{\text{el}}_j &= \sum_{\ell=1}^N E_{j} E_{\ell} 
       B_{j,\ell} \sin(\delta_{j,\ell} ) 
       ,\label{eq:active_power} \\
    Q^{\text{el}}_j &= -\sum_{\ell=1}^N E_j E_{\ell} 
      B_{j,\ell} \cos(\delta_{j,\ell}) 
       .\label{eq:reactive_power} 
\end{align}
\end{subequations}
For actual calculations, it is often convenient to use scaled units. 
In the pu system all voltages, currents, powers, and impedances are expressed in units of a suitably chosen reference value such that the load flow equations become dimensionless. 

\subsection{Fixed points of the power grid model}
\label{model_fixed_pts}

The stationary operation of a power grid corresponds to a stable fixed point of the equations of motion Eq.~\eqref{eq:inverters_pre}.
All voltages $E_j$, frequencies $\omega_j$, and phase angle differences $\delta_{j,\ell} = \delta_j - \delta_\ell$ must be constant in time to ensure a stationary power flow between the nodes of the grid.
The condition of fixed phase differences requires that all machines rotate at the same frequency $\delta_j(t) = \bar \omega t + \delta_j^\circ$ for all $j=1, \ldots, N$, leading to the conditions
\begin{equation}\label{eq:equi_point}
   \dot \omega_j = \dot E_j = 0, \quad \dot \delta_j =  \bar \omega, \quad \mbox{for all } \; j=1,\ldots,N.
\end{equation}
In a mathematically strict sense, this defines a stable limit cycle of the system.
However, all points of the cycle are physically equivalent and we can choose any point on the cycle as a representative of the equivalence class and call this an equilibrium.
We further transform to a frame of reference that rotates with a constant angular velocity of $\bar \omega$.
In this frame of reference, we have $\dot \delta_j = 0$ which simplifies the analysis.
Perturbations \textit{along} the limit cycle would add or subtract a global phase shift $\delta$ from all phases $\delta_j$ simultaneously, thus not affecting phase synchronization and power flows.
These perturbations will be excluded from the stability analysis, which is expressed in  Def.~\ref{def:stab}.
In the following, we use the superscript $\cdot^\circ$ to denote the equilibrium values of phase angle, frequency, and voltage.

Using the equations of motion Eq.~\eqref{eq:inverters_pre} and the network equations Eq.~\eqref{eq:active_power} and Eq.~\eqref{eq:reactive_power}, the equilibria of an inverter-based power grid are described by the nonlinear algebraic equations
\begin{equation} 
\begin{aligned}
   \bar \omega &= \omega_j^{\circ},  \\
   0 &= \omega^{\rm d} - \omega_j^{\circ} + \kappa_j P_j^{\rm d} - \kappa_j  \sum_{\ell=1}^N
        B_{j,\ell} E_j^{\circ} E_\ell^{\circ} 
        \sin(\delta_{j,\ell}^{\circ} ) ,
         \\
   0 &= E_j^{\rm d} - E_j^\circ + \chi_j  Q_j^{\rm d} 
     + \chi_j
      \sum_{\ell=1}^N B_{j,\ell} E_j^{\circ} E_\ell^{\circ}
       \cos(\delta_{j,\ell}^{\circ}  ) , 
\end{aligned}
\label{eq:stationary_state}
\end{equation}
in the rotating reference frame.
We note that several equilibria can coexist in networks with sufficiently complex topology, although such does not hinder performing linear stability analysis~\cite{Manik2017}. 

\section{Single Inverter Coupled to an infinite grid}
\label{sec:single_inverter}
We first examine the simplest possible system, a single inverter coupled to an infinite grid.
For this system, one can systematically compute all fixed points and scan over system parameters to obtain a comprehensive picture of the stability of the considered fixed point.

In this system, the voltage and frequency of the infinite grid are assumed to be constant at the reference level $\hat{E}$ and $\omega^d$, respectively.
In a rotating frame, we can further set the power phase angle of the grid to zero.
Hence, we are left with the dynamics of the single inverter in terms of its voltage magnitude $E_s$, power phase angle $\delta_s$, and frequency $\omega_s$.
The equations of motion Eq.~\eqref{eq:inverters_pre} then read
\begin{align}
\begin{split}
    \dot{\delta}_s &= \omega_s , \\
    \tau \dot{\omega}_s &= - \omega_s + \omega^d - \kappa (\hat{E} E_s B \sin{(\delta_s)} - P^d) ,  \\
    \tau \dot{E}_s &=  - E_s + E^d \\ 
    & \quad - \chi (-\hat{E} E_s B \cos{(\delta_s)} + B E_s^2 - Q^d ). 
\end{split}\label{eq:single_inverter}
\end{align}
where we have used that $B_{s,s} = - B_{s,\mathrm{grid}} = -B$.

\begin{figure}[t]
    \centering
    \includegraphics[width=.7\columnwidth]{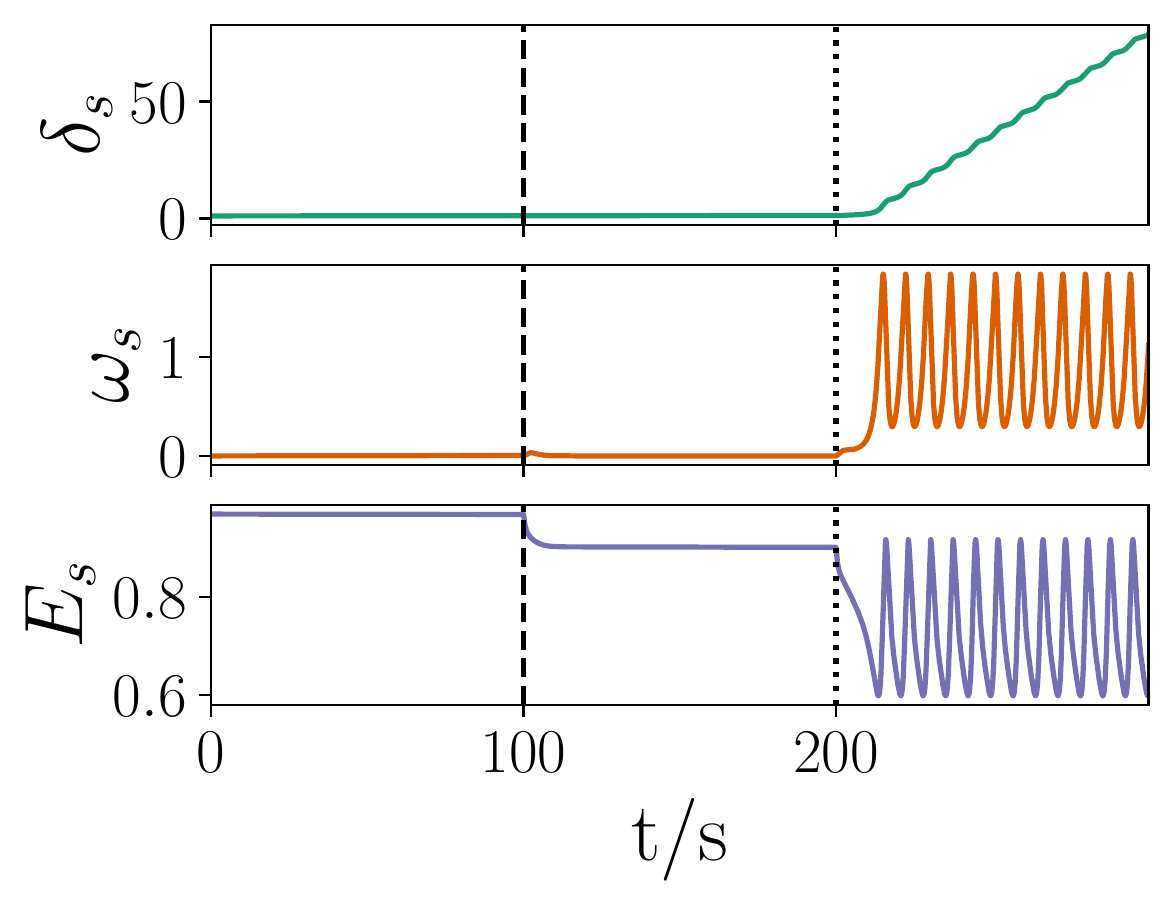}
    \caption{Simulation of a single inverter coupled to an infinite grid starting from a stable fixed point and increasing the reactive droop gain $\chi$. 
    The remaining parameters where chosen as $\tau=0.1$, $B=1.5$, $\kappa=1$, $P^d=1.5$, $Q^d=0.05$, $\hat{E}=1$, and $E^d=1$.
    Initially, the system is in a stable fixed point for a reactive droop gain of $\chi_1=0.05$. 
    After the reactive droop gain is increased to $\chi_2=0.15$ at the dashed black line, the system moves to a new stable fixed point.
    Finally, the reactive droop gain is increased to $\chi_3=0.3$ at the dotted black line. 
    After the final change to $\chi_3$, there is no stable fixed point and the dynamics exhibit a limit cycle behavior.}
    \label{fig:single_inv_example}
\end{figure}

The fixed points $\vec{x}^{\circ}= [\delta_s^{\circ}, 0, E_s^{\circ}]$ of Eq.~\eqref{eq:single_inverter} can be solved analytically by squaring the frequency and voltage equations to eliminate the phase angle $\delta_s^{\circ}$.
The fixed voltages $E_s^{\circ}$ of Eq.~\eqref{eq:single_inverter} are thus determined by the equation
\begin{align}
\begin{split}
    0 = \; &B^2 {E_s^{\circ}}^4 + 2 \chi^{-1} B {E_s^{\circ}}^3  \\
    &+ \left[ \chi^{-2} - 2 \kappa^{-1} B E^d - 2 B Q^d - \hat{E}^2 B^2\right] {E_s^{\circ}}^2  \\
    &+ \left[ -2\chi^{-2} \left( E^d + \chi Q_d \right)  \right] E_s^{\circ}  \\
    &+ \bigg[ E^d  \chi^{-2} + 2 \chi^{-1} Q^d E^d + {Q^d}^{2} + \left(\frac{\omega^d}{\kappa}\right)^2  \\ 
    &+ 2 \frac{\omega^d P^d}{\kappa} + {P^d}^{2} \bigg], 
    \end{split}
    \label{eq:single_inv_Efixed}
\end{align}
which is a 4\textsuperscript{th}-order polynomial in $E_s^{\circ}$ that can be solved analytically.
Note, only real and positive solutions of Eq.~\eqref{eq:single_inv_Efixed} are physically meaningful and will be considered in the subsequent steps.

If a solution $E_s^{\circ}$ of Eq.~\eqref{eq:single_inv_Efixed} is found, the corresponding power phase angle $\delta_s^{\circ}$ is given by
\begin{align}
    \delta_s^{\circ} &= \arcsin{\left(\frac{\omega_d / \kappa + P^d}{B \hat{E}  E_s^{\circ}}\right)}. \label{eq:single_inv_arcsin}
\end{align}
Naturally, not every set or parameter guarantees a stable fixed point.
Given a physically meaningful solution to Eq.~\eqref{eq:single_inv_arcsin}, the argument of arcsine Eq.~\eqref{eq:single_inv_arcsin} has to be in the interval $[-1, 1]$.
Note, we also presupposed that the frequency $\omega_s^\circ$ should vanish and thus we are only concerned with fixed points that meet this requirement.
We now consider an example where we see how a stable fixed point may be lost such that the inverter is unstable.
In the example shown in Fig.~\ref{fig:single_inv_example}, the simulation was initialized in a fixed point calculated using Eq.~\eqref{eq:single_inv_Efixed} and Eq.~\eqref{eq:single_inv_arcsin}. 
Here, we increase the reactive droop gain by starting with a reactive droop gain $\chi_1=0.05$, increasing it to $\chi_2=0.15$ and to $\chi_3=0.3$ at the times indicated by the vertical black lines. 
The dynamics settles on a new fixed point after the first change in $\chi$ and the stability of the fixed point is lost after the change to $\chi_3=0.3$.

\begin{figure}[t]
    \centering
    \includegraphics[width=\columnwidth]{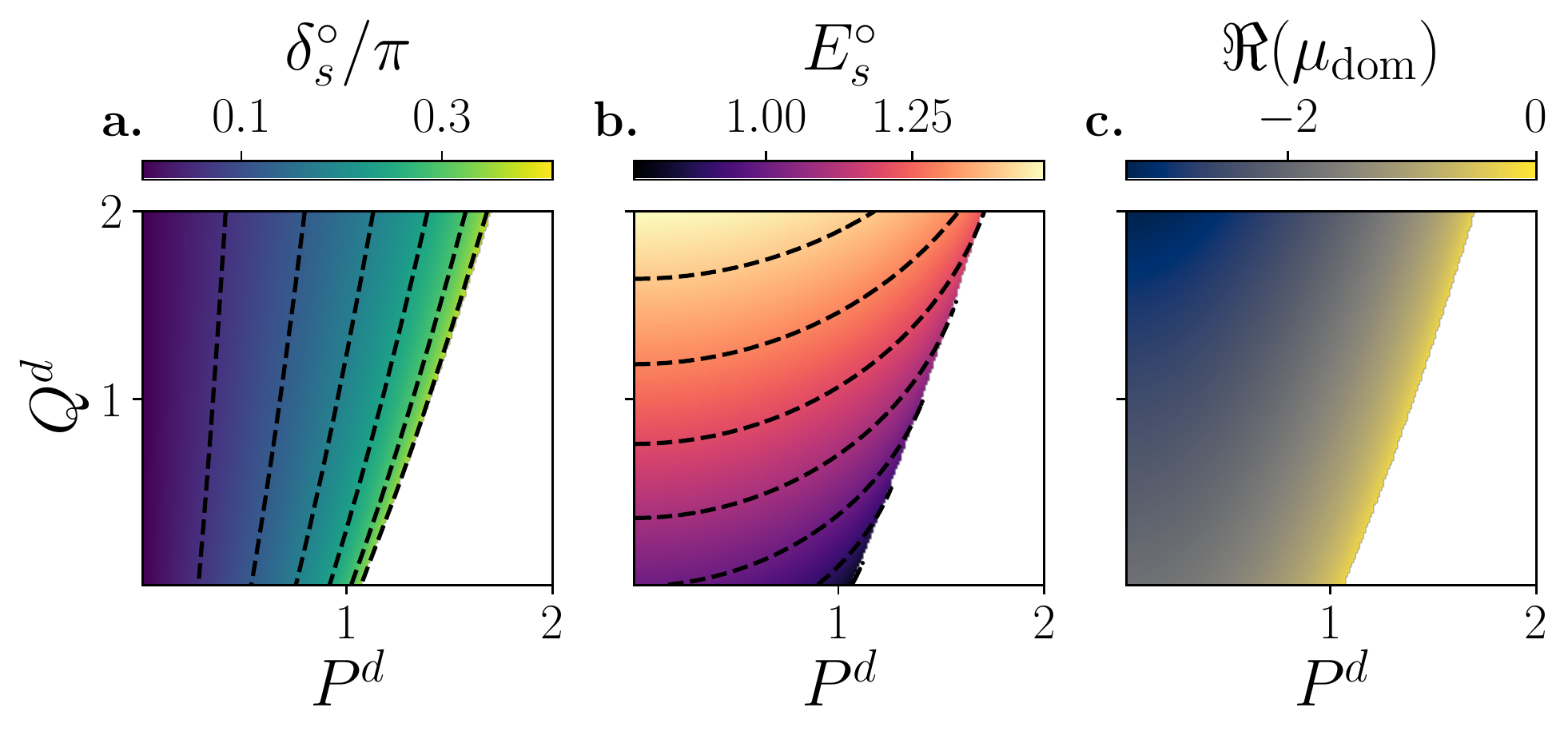}
    \caption{Scan over $Q^d$ over $P^d$ for the single inverter coupled to an infinite grid. 
    The remaining parameters were chosen as $\tau=0.1$, $B=1.5$, $\kappa=1$, $\chi=0.5$, $\hat{E}=1$, and $E^d=1$. 
    To classify the stable parameter region, the number of stable fixed points (panel \textbf{a.}), the real part of the dominant eigenvalue (panel \textbf{b.}), and the power phase angle (panel \textbf{c.}) are shown. 
    White color indicates regions where no stable fixed point could be found.
    The dashed lines are contour lines that show lines on which the fixed point power phase angle or voltage magnitude take the same value.}
    \label{fig:single_inv_QdvsPd}
\end{figure}

The local stability properties of an equilibrium, i.e., stability with respect to small perturbations, can be obtained by linearizing the equations of motion \cite{Strogatz2015, Kun04}.
We decompose the state variables into the values at the fixed point and small perturbations, which can be written as
\begin{align}
    \delta_s(t) = \delta_s^\circ + \xi_s,~
    \omega_s(t) = \omega_s^\circ + \nu_s (t),~
    E_s(t) = E_s^\circ + \epsilon_s(t).
\end{align}
We linearize  Eq.~\eqref{eq:single_inverter} in the small perturbations around the fixed point $\vec{x}^{\circ}= [\delta_s^{\circ}, 0, E_s^{\circ}]$.
The linearized dynamics are given by
\begin{align}
    \frac{d}{dt} \begin{pmatrix}
        \xi_s \\ \nu_s \\ \epsilon_s
    \end{pmatrix} = \matr{J}_s \begin{pmatrix}
        \xi_s \\ \nu_s \\ \epsilon_s
    \end{pmatrix},
\end{align}
where $\matr{J}_s \in \mathbb{R}^{3}$ is the Jacobian
\begin{equation}
\begin{aligned}\label{eq:Jabobian1}     
    &\matr{J}_s =\\ &
    \begin{pmatrix} 0 & 1 &  0 \\
    - \tau^{-1} \kappa  E_s^{\circ} C & -\tau^{-1} & - \tau^{-1}\kappa   S \\
    - \tau^{-1}\chi  E_s^{\circ} S & 0 &  -\tau^{-1}\left[1 + \chi \left(2 B E_s^{\circ} -  C\right)\right] \\
    \end{pmatrix},
\end{aligned}
\end{equation}
with $C = B \hat{E} \cos{(\delta_s^{\circ})}$ and $S = B  \hat{E} \sin{(\delta_s^{\circ})}$.

\begin{figure}[t]
    \centering\includegraphics[width=\columnwidth]{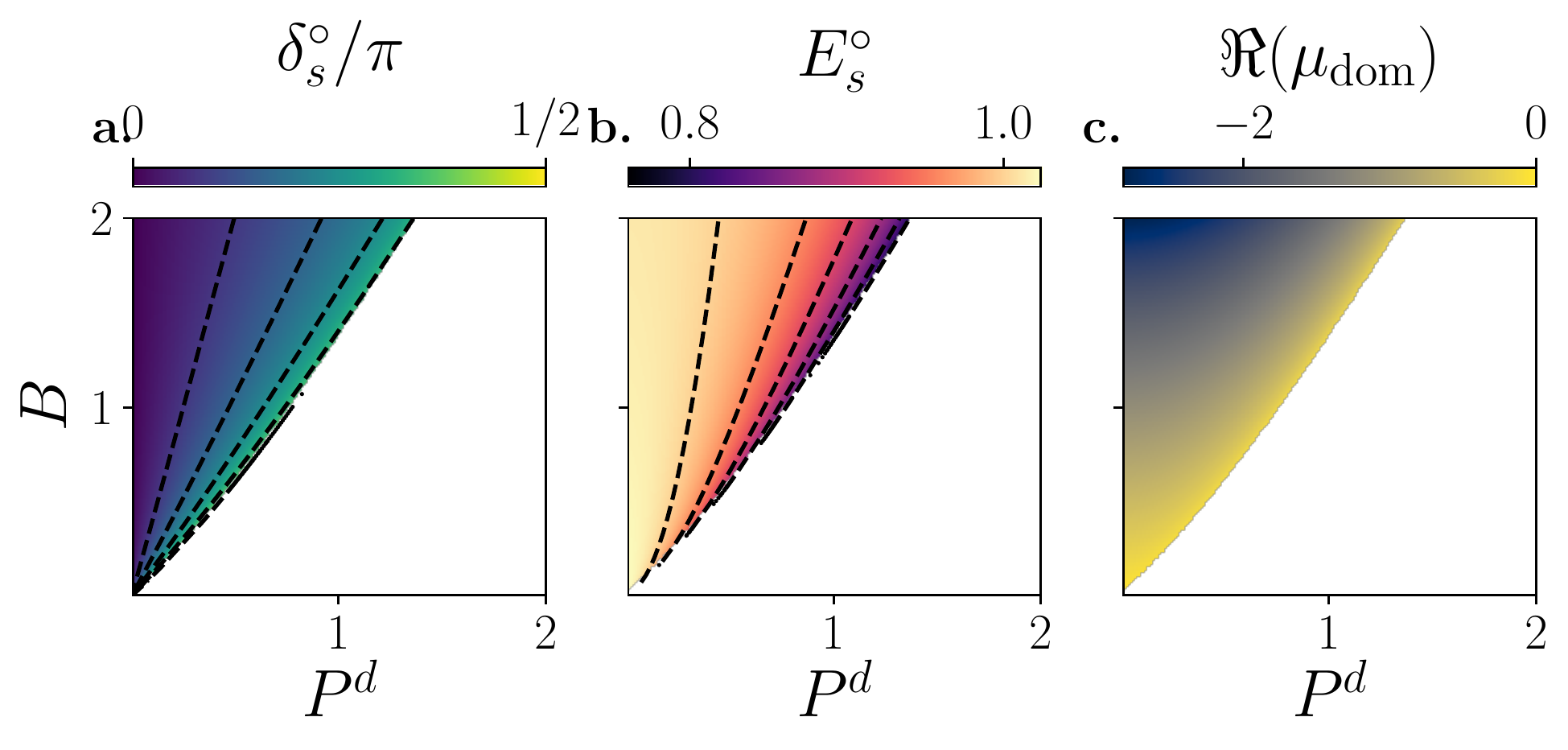}
    \caption{Scan over $B$ over $P^d$ for the single inverter coupled to an infinite grid. 
    The remaining parameters were chosen as $\tau=0.1$, $Q_d=0.05$, $\chi=0.5$, $\kappa=1$, $\hat{E}=1$, and $E^d=1$. 
    To classify how the fixed point changes for different parameters, the stationary power phase angle $\delta_a^{\circ}$ (panel \textbf{a.}), stationary voltage magnitude $E_s^{\circ}$ (panel \textbf{b.}) and the dominant eigenvalue of the Jacobian $\mu_{\mathrm{dom}}$ (panel \textbf{c.}) are recorded.
    White color indicates regions where no stable fixed point could be found.
    The dashed lines are contour lines that show lines on which the fixed point power phase angle or voltage magnitude take the same value.}
    \label{fig:single_inv_BBvsPd}
\end{figure}

The matrix $\matr {J}_s$ is the central object of linear stability analysis, with its eigenvalues and eigenvectors showing how a trajectory behaves close to a fixed point. 
The fixed point is asymptotically stable, i.e., the small disturbances $\xi_s(t)$, $\nu_s(t)$ and $\epsilon_s(t)$ decay exponentially, if all eigenvalues $\mu_i$ have a negative real part.

To understand which parameters lead to a stable fixed point, we scan over different parameter combinations.
We focus on the desired active power $P^d$, the desired reactive power $Q^d$, the coupling strength $B$, and the reactive droop gain $\chi$.
The active power $P^d$ and reactive power $Q^d$ are requirements of the connected consumers or producers of electricity, and $B$ gives the coupling strength to the power grid network.
The reactive droop gain $\chi$ is a parameter that can be freely chosen by the grid operator.
We choose $\hat{E}=1$, $E^d=1$, $\kappa=1$ and $\tau=0.1$ if not stated otherwise.
Note, while $Q^d$ can in general be negative, we focus on the case of positive $Q^d$.
In Fig.~\ref{fig:single_inv_QdvsPd} we show the static angle, static voltage, and the real part of the dominant eigenvalue examined over a range of values of $P^d$ and $Q^d$, where $\chi = 0.5$ and $B=1.5$ are fixed.

No stable fixed point can be found if the desired power exceeds a critical threshold.
The fixed point vanishes in a saddle-node bifurcation when the phase angle takes the value $\delta_s^\circ=\pi/2$ and the real part of the dominant eigenvalue $\mu_{\mathrm{dom}}$ is zero.
Increasing the desired reactive power $Q^d$ permits larger values for the desired active power $P^d$. 
Additionally, Fig.~\ref{fig:single_inv_QdvsPd} also shows that the stationary power phase angle $\delta_s^{\circ}$ is mostly influenced by the desired power $P^d$, while the stationary voltage $E_s^{\circ}$ is mostly influenced by the desired reactive power $Q^d$.

The scan over the coupling strength $B$ and desired active power $P^d$ can be seen in Fig.~\ref{fig:single_inv_BBvsPd}.
The maximum $P^d$ increases almost linearly with the coupling strength $B$ for the chosen parameters. 
Again, the stationary power phase angle $\delta_s^{\circ}$ assumes a maximum value close to the bifurcation.

\begin{figure}
    \centering
    \includegraphics[width=\columnwidth]{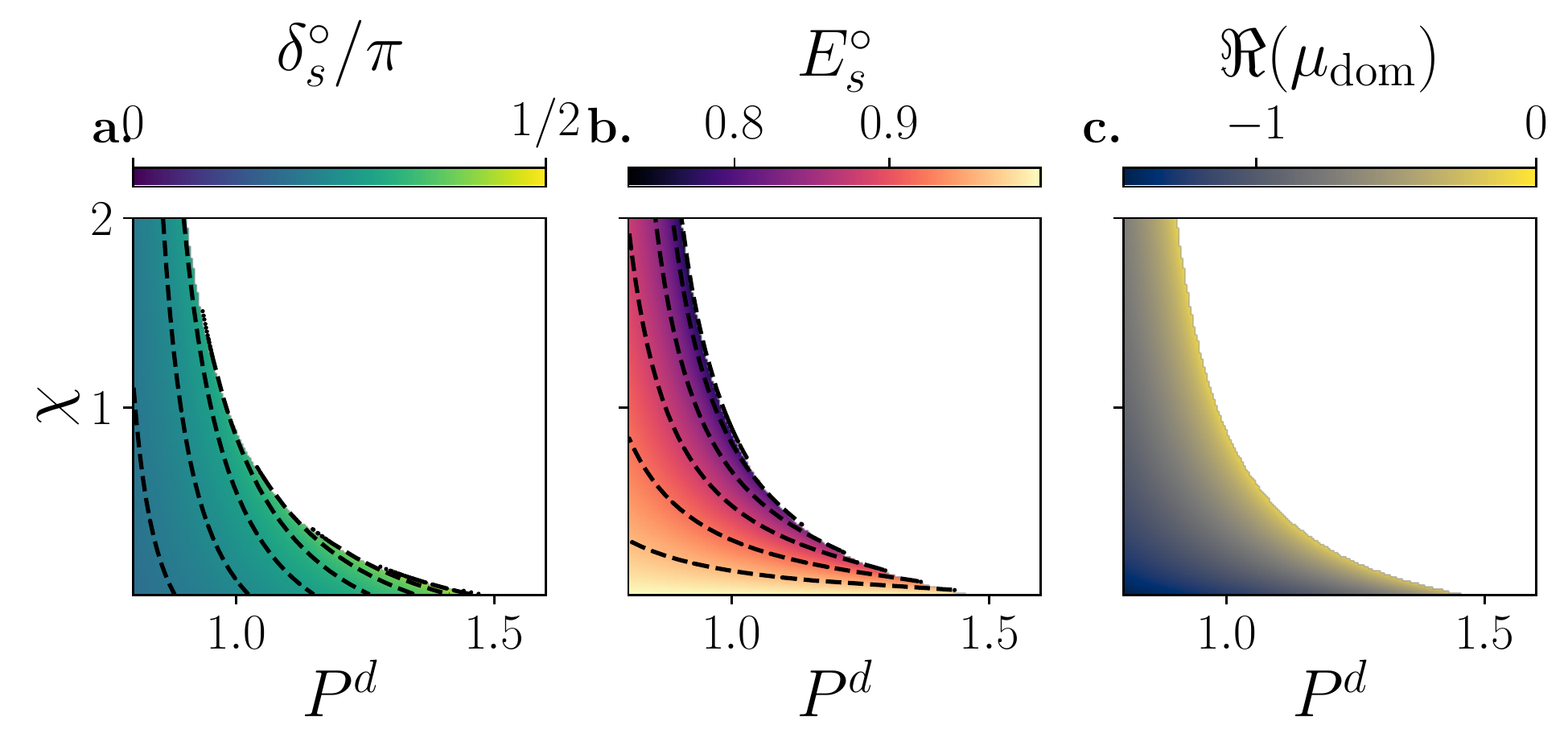}
    \caption{Scan over reactive droop gain $\chi$ and desired active power $P^d$ for the single inverter coupled to an infinite grid. 
    The remaining parameters were chosen as $\tau=0.1$, $Q_d=0.05$, $B=1.5$,  $\kappa=1$, $\hat{E}=1$, and $E^d=1$. 
    To classify how the fixed point changes for different parameters, the stationary power phase angle $\delta_a^{\circ}$ (panel \textbf{a.}), stationary voltage magnitude $E_s^{\circ}$ (panel \textbf{b.}) and the dominant eigenvalue of the Jacobian $\mu_{\mathrm{dom}}$ (panel \textbf{c.}) are recorded.
    White color indicates regions where no stable fixed point could be found. 
    The dashed lines are contour lines that show lines on which the fixed point power phase angle or voltage magnitude take the same value.}
    \label{fig:single_inv_KKvs_Pd}
\end{figure}

\begin{figure}[b]
    \centering
    \includegraphics[width=\columnwidth]{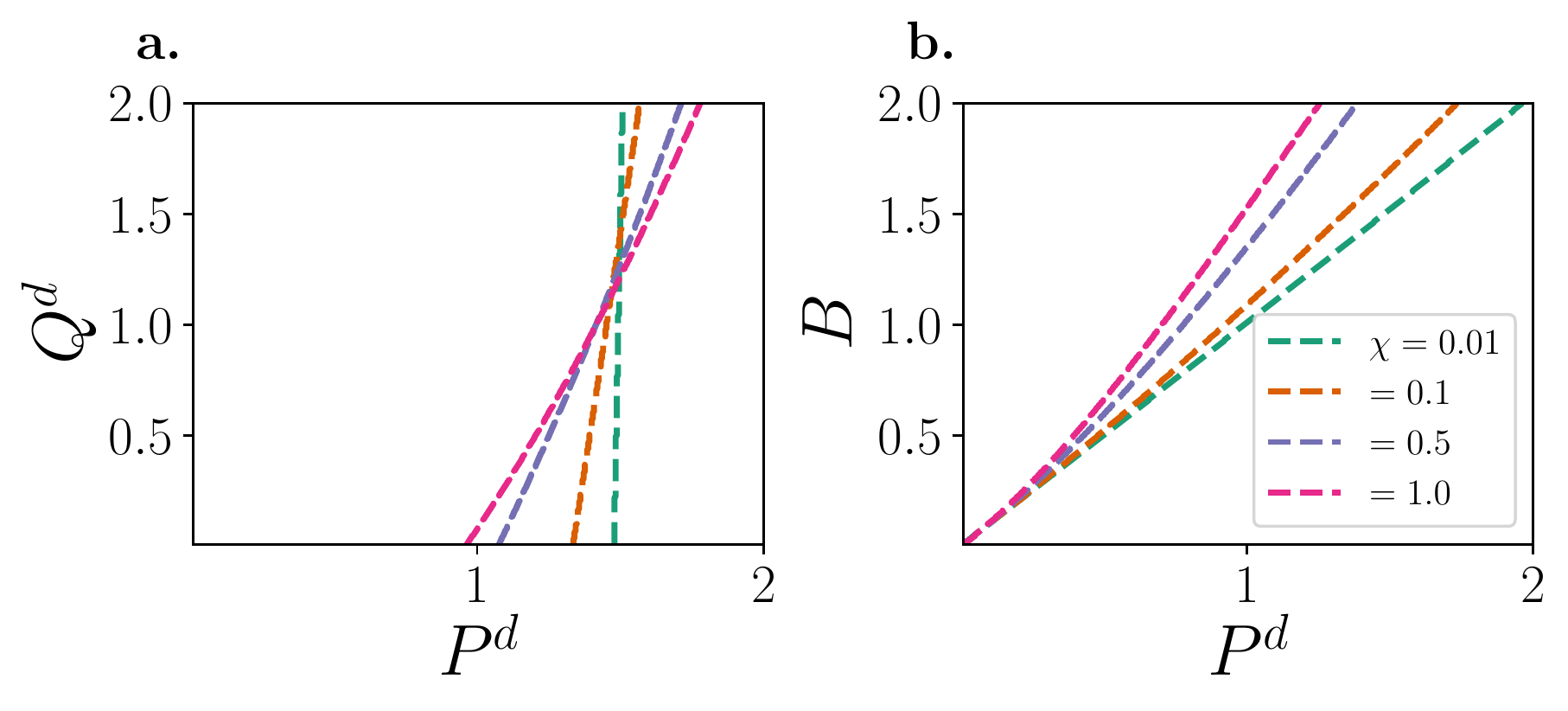}
    \caption{Shape of regions with stable fixed points for different values of the reactive droop gain $\chi$.
    Dashed lines show the border between parameter regions where a stable fixed point exists (left of the curves) and where no stable fixed point can be found (right of the curves).
    The coupling strength $B=1.5$ and the desired reactive power $Q^d=0.05$ were chosen for panel \textbf{a.} and \textbf{b.}, respectively.
    The remaining parameters were chosen as $\tau=0.1$, $\kappa=1$, $\hat{E}=1$, and $E^d=1$.}
    \label{fig:single_inv_border_vs_kkr}
\end{figure}

The scan over different reactive droop gains $\chi$ and desired power $P^d$ (see Fig.~\ref{fig:single_inv_KKvs_Pd}) reveals that there is also a bifurcation for increasing  the reactive droop gain $\chi$. 
In contrast to the parameters $B$, $P^d$ and $Q^d$, the reactive droop gain $\chi$ is a parameter that can be freely chosen in the inverters and is of large importance when trying to ensure the system stays in a stable state.

To highlight the influence of the reactive droop gain $\chi$ in shaping the parameter region with a stable fixed points, the separatrices of different values of $\chi$ are shown in Fig.~\ref{fig:single_inv_border_vs_kkr}.
For low values of $\chi$, the separatrix in the $Q^d$-$P^d$ plane is almost vertical and only $P^d$ determines if a stable fixed point is present. 
This changes for increasing values of $\chi$ resulting in the separatrix describing approximately a linear function in the $Q^d$-$P^d$ plane.
Increasing the reactive droop gain $\chi$ results in an overall smaller parameter region with a stable fixed point  in the $B$-$P^d$ plane.
Thus, the network has to be reinforced, i.e., $B$ has to be increased, to ensure that the fixed point is stable for the same $P^d$ and increasing $\chi$.

\section{Linear Stability analysis for extended networks}\label{sec:linear_stab_networks}

\subsection{Linear stability analysis}
\label{sec:linstab}
We now extend the linear stability analysis to systems consisting of multiple inverters that are coupled by an underlying network.
As in the case of the single inverter, we linearise the equations of motion around the fixed point to find out if the perturbation $\xi_j(t)$, $\epsilon_j(t)$, and $\nu_j(t)$ decay exponentially, i.e., the fixed point is linearly stable.
To this end, we decompose the state variables for each inverter $j$ as the sum of their equilibrium values and the small perturbation as 
\begin{equation}
   \delta_j(t) = \delta_j^{\circ} + \xi_j(t), \; E_j(t) = E_j^{\circ} + \epsilon_j(t), \; 
   \omega_j(t) =  \omega_j^{\circ} + \nu_j(t).  
\end{equation}
Inserting this decomposition in the equations of motion~Eq.~\eqref{eq:inverters_pre} yields, at linear order
\begin{align}
    \dot \xi_j &= \nu_j   \\
    \tau_j \dot \nu_j &= - \nu_j
    - \kappa_j \sum_{\ell=1}^N \left(
       \Lambda_{j,\ell}  \xi_\ell - A_{\ell,j} \epsilon_\ell \right) ,  \\ 
    \tau_j  \dot \epsilon_j  &= - \epsilon_j + \chi_j E_j^\circ  \sum_{\ell=1}^N \left(H_{j,\ell} \epsilon_\ell + A_{j,\ell} \xi_\ell\right), 
    \label{eq:linstab1}
\end{align}
where we have defined the matrices $\matr \Lambda, \matr A, \matr H \in \mathbb{R}^{N\!\times\!N}$ with components
\begin{align}
    \Lambda_{j,\ell} &= 
    \left\{ \begin{array}{l l}
        - E_j^{\circ} E_\ell^{\circ} B_{j,\ell} \cos(\delta_{\ell,j}^{\circ}) \; & \mbox{for} \, j \neq \ell, \\
        \sum_{k\neq j}  E_j^{\circ} E_k^{\circ} B_{j,k} \cos(\delta_{k,j}^{\circ}) \; & \mbox{for} \, j = \ell, \\
     \end{array} \right.   \\
     A_{j,\ell} &= 
    \left\{ \begin{array}{l l}
        - E_\ell^{\circ} B_{j,\ell} \sin(\delta_{\ell,j}^{\circ}) \; & \hspace*{8mm} \mbox{for} \, j \neq \ell, \\
        \sum_k E_k^{\circ} B_{j,k} \sin(\delta_{k,j}^{\circ}) \; & \hspace*{8mm} \mbox{for} \, j = \ell, \\
     \end{array} \right.     \\
    H_{j,\ell} &= 
    \left\{ \begin{array}{l l}
         B_{j,\ell} \cos(\delta_{\ell,j}^{\circ}) \; & \hspace*{0mm} \mbox{for} \, j \neq \ell, \\
        B_{j,j} + \sum_k B_{j,k} \cos(\delta_{k,j}^{\circ}) E_k^{\circ} / E_j^{\circ} \; & \hspace*{0mm} \mbox{for} \, j = \ell. \\
             \end{array} \right.   
        \label{eq:def_matrices}
\end{align}
Furthermore, we define the diagonal matrices (all in $\mathbb{R}^{N\!\times\!N}$)
\begin{align}
    \matr E &= \mathrm{diag}(E_1^\circ,E_2^\circ,\ldots,E_N^\circ),  \\
    \matr T &= \mathrm{diag}(\tau_1,\tau_2,\ldots,\tau_N),  \\
    \matr K &= \mathrm{diag}(\kappa_1, \kappa_2,\ldots,\kappa_N),  \\
    \matr X &= \mathrm{diag}(\chi_1,\chi_2,\ldots, \chi_N).  
\end{align}
In the following, we assume that all droop constants, i.e., $\kappa$ and $\chi$, and all time constants $\tau$ are strictly positive by design and we neglect the case that an equilibrium voltage vanishes exactly.
Hence, the four matrices defined above all have strictly positive diagonal entries.

We can then summarize the linearized equations of motion in a vectorial form defining the vectors $\vec \xi = (\xi_1,\ldots,\xi_N)^\top$, $\vec \nu = (\nu_1,\ldots,\nu_N)^\top$,  and $\vec \epsilon = (\epsilon_1,\ldots,\epsilon_N)^\top$, where the superscript $\top$ denotes the transpose of a matrix or vector.
We obtain
\begin{equation}\label{eq:eom-lin} 
   \frac{d}{dt} 
    \begin{pmatrix} \vec \xi \\  \vec \nu \\ \vec  \epsilon \end{pmatrix} 
    = \matr J
    \begin{pmatrix} \vec \xi \\  \vec \nu \\ \vec  \epsilon \end{pmatrix} ,
\end{equation}
with the Jacobian matrix
\begin{align}\label{eq:Jabobian2}  
    \matr J &= \begin{pmatrix} 
         \matr \eye & \matr 0 & \matr 0 \\
         \matr 0 & \matr T^{-1} \matr K & \matr 0\\
         \matr 0 & \matr 0 & \matr T^{-1} \matr X \matr E
         \end{pmatrix} 
        \begin{pmatrix} 
          \matr 0 & \matr \eye & \matr 0 \\
          -  \matr \Lambda & - \matr K^{-1}
                  &  \matr A^\top \\
          \matr A & \matr 0 & 
             \tilde{\matr{H}} \\
        \end{pmatrix} ,
\end{align}
and the abbreviation $\tilde{\matr{H}} = \matr H - \matr X^{-1} \matr E^{-1}$.

Generally, an equilibrium is linearly (asymptotically) stable if the real part of all relevant eigenvalues of the Jacobian matrix $\matr J$ is strictly smaller than zero~\cite{Strogatz2015}. 
In this case, however, we must slightly adapt this definition.
We note that $\matr J$ always has one eigenvalue $\mu_1 = 0$ with the eigenvector
\begin{equation}
    \begin{pmatrix} \vec \xi &  \vec \nu & \vec \epsilon \end{pmatrix}^\top
     = \begin{pmatrix} \vec 1 & \vec 0 & \vec 0 \end{pmatrix}^\top\!.
\end{equation}
This eigenvector corresponds to a global shift of the inverters' phase angles $\vec{\delta} \mapsto \vec{\delta} + \alpha$ which has no  physical significance and must thus be excluded from the stability analysis.
Hence, the linear stability analysis can be restricted to the subspaces 
\begin{subequations}\label{eq:Subspaces}
\begin{align}
   \mathcal{D}^{(3)}_\perp &= \left\{ (\vec \xi,\vec{\nu}, \vec \epsilon)\in \mathbb{R}^{3N}|(\vec{1, 0, 0})^\top(\vec \xi, \vec \nu, \vec \epsilon) = 0\right\},\\
   \mathcal{D}^{(2)}_\perp &= \left\{ (\vec \xi, \vec \epsilon) \in \mathbb{R}^{2N}|(\vec{1, 0})^\top(\vec{\xi}, \vec{\epsilon})=0\right\},\\
   \mathcal{D}^{(1)}_\perp &= \left\{ \vec \xi \in \mathbb{R}^{N}|\vec{1}^\top\vec{\xi}=0\right\} .
  \end{align}
\end{subequations}
Furthermore, it is convenient to order the eigenvalues of the Jacobian as
\begin{equation}
   \mu_1=0, ~~ \Re (\mu_2) \le  \Re (\mu_3) \le  \cdots \le  \Re (\mu_{3N}) .
\end{equation}
We can now formulate a consistent condition for linear (asymptotic) stability transversally to the limit cycle (cf. Ref.~\cite{Strogatz2015}).
\begin{defn}\label{def:stab}
The equilibrium $(\delta^\circ_j, \omega^\circ_j, E^\circ_j)$ is linearly (asymptotically) stable if $\Re(\mu_n) < 0$  for all eigenvalues $n = 2,\ldots,3N$ of the Jacobian matrix $\matr J$ defined in Eq.~\eqref{eq:Jabobian2}. 
\end{defn}

\subsection{The reduced Jacobian}
\label{sec:raduced-jacobian}

We can significantly simplify the linear stability analysis by eliminating the frequency subspace, leading to a reduced Jacobian of dimension $2N$ instead of $3N$. 
In particular, we obtain the following lemma
\begin{lemma}\label{lemma:Xi}
The linear stability of an equilibrium $(\delta^\circ_j, \omega^\circ_j, E^\circ_j)$ is determined by the reduced Jacobian
\begin{equation}
   \matr \Xi = 
    \begin{pmatrix}
    - \matr \Lambda & \matr A^\top \\
    \matr A &  \tilde{\matr H} 
    \end{pmatrix}.
    \label{eq:def-Xi-lossless}
\end{equation}
The equilibrium is stable if $\matr \Xi$ is negative definite on ${\mathcal{D}}_\perp^{(2)}$. It is unstable if $\matr \Xi$ is not negative semi-definite on ${\mathcal{D}}_\perp^{(2)}$.
\end{lemma}
\begin{proof}
Define the Lyapunov function candidate
\begin{equation}
\begin{aligned}
  V := 
  \begin{pmatrix} 
     \vec \nu \\ \vec \xi \\ \vec \epsilon 
  \end{pmatrix}^{\!\!\top}
  \underbrace{\begin{pmatrix} 
      \matr K^{-1} \matr T & 0 & 0 \\ 
      0 & \matr \Lambda  & -\matr A^\top \\
      0 & -\matr A & -\tilde{\matr H}
  \end{pmatrix}}_{=: \matr P} \, 
  \begin{pmatrix} 
      \vec \nu \\ \vec \xi \\ \vec \epsilon 
  \end{pmatrix} \, .
\end{aligned}
\end{equation}
Then one finds
\begin{equation}
\begin{aligned}
  \dot V &= -2 \vec \nu^\top \matr  K^{-1}\vec \nu
      - 2 (\matr A \vec \xi + \tilde{\matr H} \vec \epsilon)^\top
        \matr T^{-1} \matr X \matr E 
      (\matr A \vec \xi + \tilde{\matr H} \vec \epsilon) \\
     &\le 0. 
\end{aligned}
\end{equation}
The last inequality follows as the matrices $\matr T, \matr X$, $\matr K$, and $\matr E$ are diagonal with only positive entries. 
If $\matr \Xi$ is negative definite, then $\matr P$ is positive definite and the equilibrium is stable according to the Lyapunov stability theorem.
If $\matr \Xi$ is not negative semi-definite, then also $\matr P$ is not positive semi-definite and the equilibrium is unstable according to the Lyapunov instability theorem.
\end{proof}

\section{Explicit conditions for stability and instability}
\label{sec:explicit_conditions}

In this section, we derive some explicit conditions for the stability or instability of an inverter-based grid.
Guided by the results for the single inverter from Sec.~\ref{sec:single_inverter}, we will focus on the role of the network connectivity and the reactive droop gain $\chi$.
The starting point of our analysis is Lemma~\ref{lemma:Xi}, for which we introduce a further decomposition in the voltage and angle subspace.

\subsection{Decomposing voltage and angle subspaces}

We can obtain further insight into the stability condition by a decomposition in terms of the rotor angle and the voltage dynamics.
Applying the Schur or Albert complement~\cite{zhang2006schur} to the reduced Jacobian $\matr \Xi$ we obtain the following result.

\begin{lemma}[Sufficient and necessary stability conditions for lossless power grids]\label{Prop:Schur}\
\begin{itemize}
\item[I.] The equilibrium $(\delta^\circ_j, \omega^\circ_j, E^\circ_j)$ is linearly stable if 
(a) the matrix $\matr \Lambda$ is positive definite on $\mathcal{D}^{(1)}_\perp$ and
(b) the matrix $\tilde{\matr H} + \matr A \matr \Lambda^{+} \matr A^\top$ is negative definite, where $\cdot^{+}$ is the Moore--Penrose pseudoinverse.
\item[II.] The equilibrium $(\delta^\circ_j, \omega^\circ_j, E^\circ_j)$ is linearly stable if 
(a) the matrix $\tilde{\matr H}$ is negative definite and
(b) the matrix $\matr \Lambda  + \matr A^\top \tilde{\matr H}^{-1}\matr A $ is positive definite on $\mathcal{D}^{(1)}_\perp$.
\end{itemize}
The equilibrium is linearly unstable if any of the above definiteness conditions are violated.
\end{lemma}

\begin{proof}
Here, we present solely the proof for criterion~I, as an equivalent procedure can be used to prove criterion~II.
The reduced Jacobian matrix $\matr{\Xi}$ can be decomposed as
\begin{equation}\label{eq:Schur}
  \matr{\Xi} = \matr{U}^\top\matr{S}\matr{U},
\end{equation}
with $\matr{S}\in\mathbb{R}^{2 N\times 2N}$ a block diagonal matrix
\begin{equation}
\matr S = \begin{pmatrix} 
  -\matr \Lambda  & \matr 0  \\
  \matr 0  & \tilde{\matr H} + \matr A \matr \Lambda^{+} \matr A^\top 
\end{pmatrix},
\end{equation}
and a transformation matrix
\begin{equation}
   \label{eq:defU}
   \matr U = 
   \begin{pmatrix} 
     \eye & - \matr \Lambda^{+} \matr{A}^\top \\ 
     0 & \eye 
     \end{pmatrix}.
\end{equation}
The transformation matrix $\matr U$ is of full rank and maps the vector $(\vec 1, \vec 0)^\top$ onto itself. 
Hence, $\matr U$ also maps the relevant subspace $\mathcal{D}^{(2)}_\perp$ onto itself. 
Now assume that $\matr S$ is negative definite on $\mathcal{D}^{(2)}_\perp$. Then for every $\vec x \in \mathcal{D}^{(2)}_\perp$, $\vec x \neq \vec 0$ we have
\begin{equation}
    \vec x^\top \matr \Xi \vec x =
    (\matr U \vec x)^\top \matr S (\matr U \vec x) < 0. 
\end{equation}
Similarly, if we assume that $\matr \Xi$ is negative definite on $\mathcal{D}^{(2)}_\perp$, then for every $\vec y \in \mathcal{D}^{(2)}_\perp$, $\vec y \neq \vec 0$ we have
\begin{equation}
    \vec y^\top \matr S \vec y =
    (\matr U^{-1} \vec y)^\top \matr \Xi  (\matr U^{-1} \vec y) < 0.
\end{equation}
Hence, the transformation $\matr U$ does not affect the definiteness: $\matr \Xi$ is negative (semi-)definite on $\mathcal{D}^{(2)}_\perp$ if and only if $\matr S$ is negative (semi-)definite on $\mathcal{D}^{(2)}_\perp$.

Using Lemma~\ref{lemma:Xi} we know that the equilibrium is linearly stable if $\matr \Xi$ or equivalently $\matr S$ is negative definite on $\mathcal{D}^{(2)}_\perp$ and unstable if $\matr \Xi$ is not negative semi-definite.
Since $\matr S$ is block diagonal, the definiteness of the entire matrix is equivalent to the definiteness of both blocks and the lemma follows. 
\end{proof}

Lemma~\ref{Prop:Schur} allows us to obtain a deeper insight into the linear stability of inverter-based power grids and permits the derivation of several explicit stability criteria.
To this end, we will proceed through the conditions of the lemma step by step.

\subsection{Angle Stability}\label{sec:lossless-angle}

Condition~I. (a) in Lemma~\ref{Prop:Schur} refers to the stability of the isolated phase angle system, disregarding the voltage dynamics.
To see this, we artificially fix the voltages such that $\epsilon = 0$.
The linearised equations of motion read
\begin{equation}
  \frac{\mathrm{d}}{\mathrm{d} t}
  \begin{pmatrix}
    \vec{\xi}\\ \vec{\nu}
  \end{pmatrix} =
  \begin{pmatrix} \matr 0 & \matr \eye \\
      - \matr T^{-1} \matr K \matr \Lambda 
      & - \matr T^{-1} 
   \end{pmatrix}
  \begin{pmatrix}
      \vec{\xi}\\ \vec{\nu} 
   \end{pmatrix}.
\end{equation}
Performing the same simplification as in the previous sections, one finds that the system is stable if and only if the matrix $\matr \Lambda$ is positive definite on $\mathcal{D}^{(1)}_\perp$, which is identical to condition~I.(a) in Lemma~\ref{Prop:Schur}. 
One can now easily find a sufficient condition for angle stability.
If for all connections $(j,\ell)$ in the grid we have
\begin{equation}\label{eq:limit_angles}
   \cos(\delta_j^{\circ} - \delta_\ell^{\circ}) > 0,
\end{equation}
then $\matr \Lambda$ is a proper Laplacian matrix of a weighted undirected network, which is well known to be positive definite on $\mathcal{D}_\perp^{(1)}$~\cite{Newman2018}. 

Necessary and sufficient conditions are harder to obtain. If condition Eq.~\eqref{eq:limit_angles} is not satisfied for one or several connections, the matrix $\matr \Lambda$ rather describes a signed graph, for which positive definiteness is more involved. 
A variety of criteria have been obtained in~\cite{Chen20161, Chen20162, Song2015, Zelazo2014}. 

In the following derivations, the Fiedler value or algebraic connectivity $\lambda_2$ and the corresponding Fiedler vector $v_F$ will be used~\cite{Fiedler1973, Fiedler1975, Chung1997}.
They are the smallest non-zero eigenvalue $\lambda_2$ of a Laplacian and the corresponding eigenvector $v_F$.
A Laplacian has at least one zero-valued eigenvalue $\lambda_1=0$.
Further zero-valued eigenvalues correspond to the number of disjoint connected components in the network.
The algebraic connectivity $\lambda_2$, as its name suggests, indicates how well-connected a network is. 
As the connection between different parts of the networks become weaker, $\lambda_2$ moves closer and closer to zero until the network separates into disjoint parts increasing the multiplicity of $\lambda_1$.
Thus, the algebraic connectivity $\lambda_2$ encodes relevant information regarding the connectivity of the network and plays a crucial role in the stability analysis of networked systems.

\subsection{Voltage Stability}
\label{sec:lossless-voltage}

Condition~II. (a) in Lemma~\ref{Prop:Schur} entails the stability of the isolated voltage subsystem. 
To see this, we artificially fix the angles such that $\vec \nu = \vec \xi  = \vec 0$.
Then the linearized equations of motion read
\begin{align}
    \frac{\mathrm{d}}{\mathrm{d} t} \vec \epsilon =
     \matr{T}^{-1} \matr X \matr E \tilde{\matr H} \,\vec \epsilon .
\end{align}
Recall that the matrix $\matr{T}^{-1} \matr X \matr E$ is a diagonal matrix with strictly positive entries.
Hence, we find that the isolated voltage dynamics is linearly stable if and only if the matrix $\tilde{\matr H}$ is negative definite, which is identical to condition~II. (a) in Lemma~\ref{Prop:Schur}. 

We find a necessary and a sufficient condition for voltage stability in terms of the droop constants (cf.~\cite{Sharafutdinov2018}).
Both show that the droop gains for the voltage control must not be chosen too large.
\begin{corr}\label{cor:voltage1}
If, for all nodes $j=1,\ldots,N$,
\begin{equation}
    \frac{1}{\chi_j} >  \sum_{\ell =1}^N B_{j,\ell} (E_j^{\circ} + E_\ell^{\circ})
\end{equation}
and condition Eq.~\eqref{eq:limit_angles} holds for all connections $(j,\ell)$ in a power grid, then the matrix $\tilde{\matr H}$ is negative definite and the voltage sub-system is stable.
\end{corr}
\begin{proof}
By applying Ger\v{s}gorin's circle theorem~\cite{Gerschgorin1931} to the matrix $\tilde{\matr H}$, the following condition for its eigenvalues $\alpha_j, \forall j$ stands
\begin{equation}
    |\alpha_{j} -\mathcal{C}_j| \le R_j ,
\end{equation}
where
\begin{equation}
\begin{aligned}
    \mathcal{C}_j &= \tilde{H}_{j,j} = H_{j,j} - (\matr E^{-1} \matr X^{-1})_{j,j} \\
    &= B_{j,j} + \sum_k B_{j,k} \cos(\delta_k^{\circ} - \delta_j^{\circ}) \frac{E_k^{\circ} }{E_j^{\circ}} 
    - \frac{1}{\chi_j E_j^{\circ}} \\
    R_j &= \sum_{\ell \neq j}^N |B_{j,\ell} \cos{(\delta_{j}^{\circ} - \delta_{l}^{\circ})}| ,
\end{aligned}
\end{equation}
are the center and the radius of the Ger\v{s}gorin disk, respectively.
All eigenvalues are guaranteed to lie in the left half of the complex plane, which yields the negative definiteness, if $\mathcal{C}_j + R_j < 0$ for all $j=1,\ldots, N$. Evaluating this condition yields
\begin{equation}
\begin{aligned}
   \frac{1}{\chi_j} > \; &
   B_{j,j} E_j^{\circ}  + \sum_k B_{j,k} \cos(\delta_k^{\circ} - \delta_j^{\circ}) E_k^{\circ} \\
  & \qquad + \sum_{\ell \neq j}^N  E_j^{\circ} | B_{j,\ell}  \cos{(\delta_{j}^{\circ} - \delta_{\ell}^{\circ})} | .
\end{aligned}
\end{equation}
Using the bound $\cos(\cdot) \le 1$, $|\cos(\cdot)| \le 1$, and the positivity of $B_{j,\ell}$, a sufficient condition for negative definiteness of $\tilde{\matr{H}}$ is obtained as
\begin{equation}
    \frac{1}{\chi_j} >  \sum_{\ell =1}^N B_{j,\ell} (E_j^{\circ} + E_\ell^{\circ}) \, .
\end{equation}
This concludes the proof. \end{proof}

\begin{corr}\label{cor:voltage2}
If for any subset of nodes $\mathcal{S} \subseteq \{1,2,\ldots,N\}$,
\begin{equation}
    \sum_{j \in \mathcal{S}} 
      \frac{1}{\chi_j E_j^{\circ}}
    \le \sum_{j,\ell \in \mathcal{S}} H_{j,\ell},
\end{equation}
then the matrix $\tilde{\matr H}$ is not negative definite and the equilibrium is linearly unstable.
\end{corr}
\begin{proof}
This result follows from evaluating the expression $\vec x^\top \tilde{\matr H} \vec x$ for a trial vector $\vec x \in \mathbb{R}^N$ with entries $x_j = 1\; \forall j \in \mathcal{S}$ and $x_j = 0\; \forall j \notin \mathcal{S}$. 
\end{proof}

\subsection{Coupled stability criteria}\label{sec:lossless-coupled}

We have shown that part (a) of both stability criteria in Lemma~\ref{Prop:Schur} refer to each isolated subsystem.
Hence both must be stable in themselves to enable linear stability of the full dynamical system.
The remaining parts (b) of the stability criteria then refer to the coupled frequency and voltage dynamics.
These criteria are significantly stricter than the isolated criteria.
To see this we focus on criterion I in Lemma~\ref{Prop:Schur}, assuming that $\matr \Lambda$ is positive definite on $\mathcal{D}^{(1)}_\perp$ is stable.
The complementary conditions~I. (b) is
\begin{equation}\label{eq:conIIb-succ}
   \tilde{\matr H} + \matr A  \matr \Lambda^{+} \matr A^\top \prec 0,
\end{equation}
where $\matr Z \prec 0$ is used as a short hand for $\matr Z$ being negative definite.
This condition is stricter than the condition of pure voltage stability, $\tilde{\matr H} \prec 0$.
Hence, the stability of the two isolated subsystems is not sufficient. 
Instead, they must comprise a certain `security margin' quantified by the second term on the left-hand side of Eq.~\eqref{eq:conIIb-succ}, in order to maintain linear stability.

We now present several explicit stability criteria, focusing on the interplay of internal dynamics and the grid topology.
We typically assume that the two isolated subsystems are stable, i.e., conditions~I. (a) and II. (a) in Lemma~\ref{Prop:Schur} are satisfied unless stated otherwise.
First, we consider the case of small voltage droop gains, as this is necessary to ensure voltage stability, see Corollaries~\ref{cor:voltage1} and \ref{cor:voltage2}, and relate stability to the connectivity of the grid.

\begin{corr}\label{cor:voltage_stability_graph}
A necessary condition for the stability of an equilibrium point is given by
\begin{equation}\label{eq:stab-lambda-small-x}
    \lambda_2 > \sum_{j=1}^N
    \chi_j E_j^\circ
    \left[ \sum_{k=1}^N A_{j,k} v_{Fk} \right]^2
    + \mathcal{O}(\chi_j^2),
\end{equation}
where $\lambda_2$ is the network's algebraic connectivity and $\vec v_F$ denotes the Fiedler vector of the Laplacian $\matr \Lambda$ for $\chi_j \equiv 0$.
\end{corr}
\begin{proof}
A normalized Fiedler vector $\vec v_F$ is defined at $\chi_j \equiv 0$~\cite{Fiedler1973, Fiedler1975, Chung1997}. 
The actual normalized Fiedler vector, for a particular non-zero value of the $\chi_j$, is denoted $\vec v'_F$, such that
\begin{equation}
   \vec v'_F = \vec v_F + 
   \mathcal{O}(\chi_j). 
\end{equation}
Following criterion II.~(b) in Lemma~\ref{Prop:Schur}, the stability of the fixed point requires that all vectors $\vec y$ obey
\begin{equation}
    \vec y^\top \matr \Lambda \vec y > -
    \vec y^\top \matr A^\top 
    \tilde{\matr H}^{-1} \matr A \vec y.
\end{equation}
Now we consider the particular choice $\vec y = \vec v'_F$ to obtain a necessary condition for stability. 
Furthermore, we expand the matrix inverse according to
\begin{equation}
   -\tilde{\matr H}^{-1} = (\matr E^{-1} \matr X^{-1} - \matr H)^{-1} =  \sum_{\ell = 0}^\infty \matr E \matr X
   (\matr E \matr X \matr H)^\ell,
\end{equation}
up to the leading term. 
To leading order in the voltage droop gains, we then obtain the necessary condition
\begin{equation}
   \lambda_2 >
   \vec v_F^\top \matr A^\top \matr E \matr X
   \matr A \vec v_F + \mathcal{O}(\chi_j^2). 
\end{equation}
An equivalent result in synchronous generators can be found in Ref.~\cite{Sharafutdinov2018} and concludes the proof.
\end{proof}

We recall that a necessary and sufficient criterion for the stability of the isolated frequency subsystem is given by $\lambda_2 > 0$. 
In contrast, the right-hand side of condition Eq.~\eqref{eq:stab-lambda-small-x} is generally positive.
Hence, additional algebraic connectivity is needed in the grid as a `security margin' to guarantee the stability of the whole dynamical system.

We can further derive two sufficient stability criteria, one in terms of the reactive droop gains $\chi$ by extending Corollary~\ref{cor:voltage1} and one in terms of the algebraic connectivity $\lambda_2$ of the network.

\begin{corr}\label{cor:4}
An equilibrium is linearly stable if the network's algebraic connectivity is positive, $\lambda_2>0$, and the reactive droop gains satisfy
\begin{equation}\label{eq:stab-cond-Xl}
   \frac{1}{\chi_j } 
     > \sum_{\ell=1}^N B_{j,\ell} + E_j^\circ
     \frac{\| \matr A  \|_{2}  \| \matr A^\top \|_{2}  }{\lambda_2},
\end{equation}
for all nodes $j=1,\ldots, N$, where $\| \cdot \|_2$ is the induced $\ell_2$-norm.
\end{corr}

\begin{proof}
A positive algebraic connectivity $\lambda_2 > 0$ implies that both $\matr \Lambda$ is positive definite on $\mathcal{D}^{(1)}_\perp$ and criterion I.~(a) in Lemma~\ref{Prop:Schur} is satisfied.

Consider now criterion I.~(b). Using the same arguments as in the proof of Corollary~\ref{cor:voltage1} one can show that the conditions Eq.~\eqref{eq:stab-cond-Xl} imply that the matrix
\begin{equation}
   \tilde{\matr H} + \frac{\| \matr A  \|_{2}  
      \| \matr A^\top \|_{2}}{\lambda_2 }  \; 
      \eye 
\end{equation}
is negative definite. 
Noting that $\lambda_2^{-1} = \| \matr \Lambda^+ \|_{2}$ and using the sub-multiplicativity of the $\ell_2$-norm, we then find that $\forall \vec y \neq \vec 0$ we have
\begin{equation}
\begin{aligned}
    & \vec y^\top  \left[ 
    -\tilde{\matr H}
    -  \frac{\| \matr A  \|_{2}  
      \| \matr A^\top \|_{2}}{\lambda_2 }
    \right] \vec y > 0  \\
    \Leftrightarrow \; &
    \vec y^\top  \left[ 
   \matr E^{-1} \matr X^{-1}- \matr H
    \right] \vec y >
    \| \matr A \|_{2} \| \matr \Lambda^+ \|_{2} \| \matr A^\top \|_{2} \| \vec y \|^2  \\
    & \qquad \qquad \ge \| \matr A  \matr \Lambda^+ \matr A^\top \|_{2} 
        \| \vec y \|^2  \\
    & \qquad \qquad  \ge \vec y^\top \matr A \matr \Lambda^+ \matr A^\top \vec y \\
     \Leftrightarrow \; &
    \vec y^\top  \left[ 
    \matr E^{-1} \matr X^{-1} - \matr H
    -  \matr A \matr \Lambda^+ \matr A^\top
    \right] \vec y > 0 .
\end{aligned}
\end{equation}
Hence, matrix $\tilde{\matr H} + \matr A  \matr \Lambda^+ \matr A^\top$ is negative definite. Criterion I. (b) in Lemma~\ref{Prop:Schur} is satisfied and the equilibrium is linearly stable.
\end{proof}

The condition given by Corollary~\ref{cor:4} highlights that a stable fixed point needs both a sufficiently large algebraic connectivity $\lambda_2$ and low reactive droop gains $\chi$.

\begin{corr}\label{cor:5}
If by criterion II.~(a) in Lemma~\ref{Prop:Schur} the matrix $\tilde{\matr H}$ is negative definite, and if the algebraic connectivity $\lambda_2$ satisfies
\begin{equation}\label{eq:stab-cond-lam}
    \lambda_2 > \| \matr A^\top \tilde{\matr H}^{-1}  \matr A \|_2,
\end{equation}
where $\| \cdot \|_2$ is the induced $\ell_2$-norm, then the equilibrium point is linearly stable.
\end{corr}

\begin{proof} 
Condition II.~(a) in Lemma~\ref{Prop:Schur} is satisfied by assumption, so we can focus on condition II.~(b)

The assumption Eq.~\eqref{eq:stab-cond-lam} implies that $ \forall \vec y \in \mathcal{D}^{(1)}_\perp$
\begin{equation}
\begin{aligned}
   \vec y^\top  \matr \Lambda  \vec y 
      & \ge \lambda_2 \|\vec y\|^2  \\
   & > \| \matr A^\top  
      \tilde{\matr H}^{-1} 
      \matr A \|_2 \|\vec y\|^2  \\
   & \ge \vec y^\top \matr A^\top  \tilde{\matr H}^{-1} \matr A \vec y, 
\end{aligned}
\end{equation}
such that the matrix $\matr \Lambda + \matr A^\top \tilde{\matr H}^{-1} \matr A$ is positive definite on $\mathcal{D}^{(1)}_\perp$. 
Condition II.~(b) in Lemma~\ref{Prop:Schur} is therefore satisfied and the equilibrium is linearly stable.
\end{proof}

\section{Testing the Usefulness of the Stability Criteria}\label{sec:numerical_results}

\begin{figure}
    \centering   \includegraphics[width=\columnwidth]{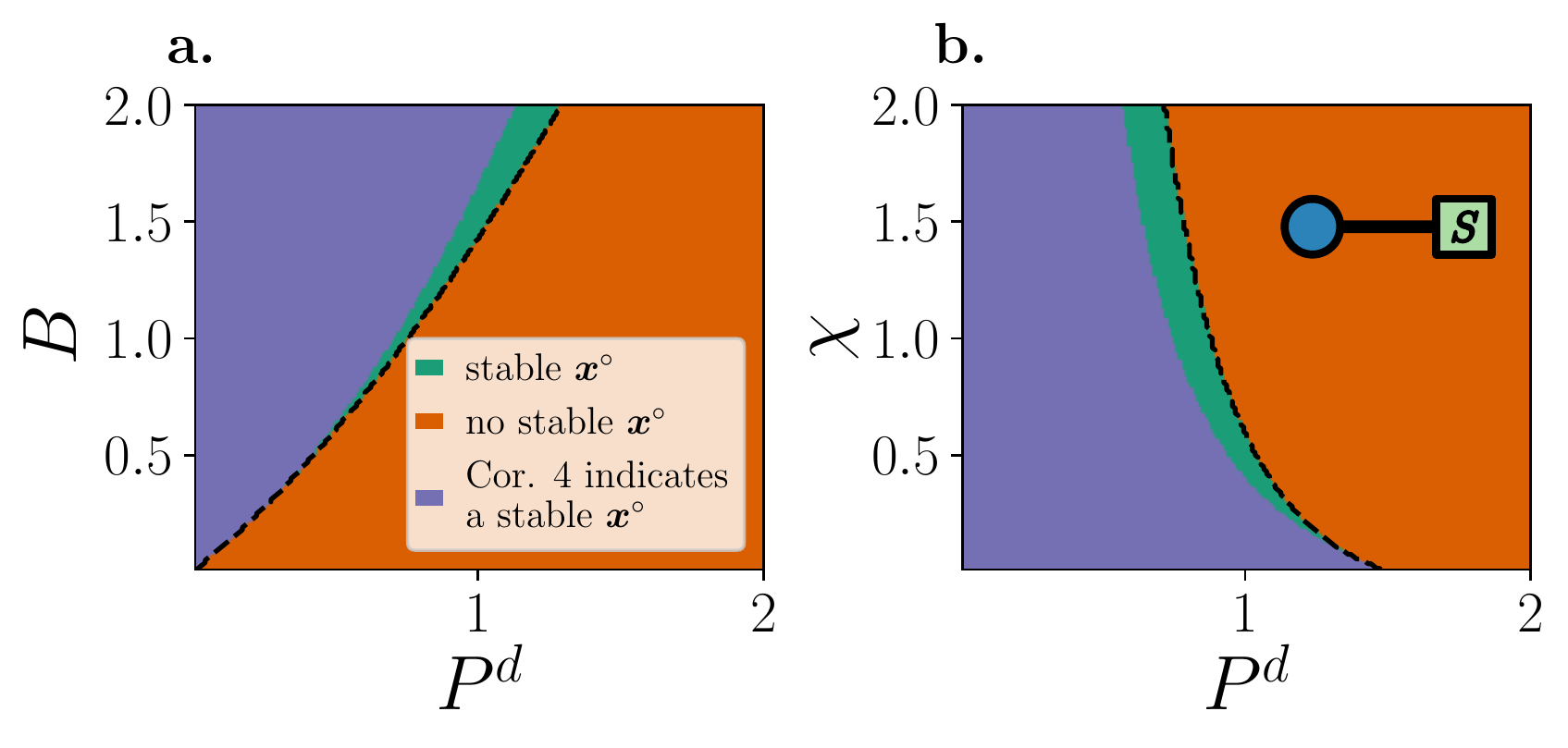}\caption{Corollary~\ref{cor:4}, a sufficient condition for stability, results in a tight bound for the region with a stable fixed point for a two-inverter system.
    A schematic representation of the network is shown in the inset in the upper right of panel \textbf{b.}, with one of the inverters acting as a slack node (indicated by $S$).
    Panels \textbf{a.} and \textbf{b.} are the results of a scan over $P^d$ and, respectively, the coupling strength $B$ and reactive droop gain $\chi$. 
    The different colors show parameter regions with a stable fixed point, no-stable fixed points, and where Corollary~\ref{cor:4} implies the fixed point is stable.
    The separatrix between stable and no-stable fixed points regions is given by the dashed black line.
    Note that the region where the Corollary~\ref{cor:4} implies the stability of the fixed point overlays the region with the numerically stable fixed point.}
    \label{fig:two_inverter_stable_map}
\end{figure}

We now compare the findings above to numerical results for two test systems.
One system consists of two inverters and the other consists of ten inverters arranged in a tree-like topology. 
In both cases, a subset of the inverters act as net producers of electricity, which can be seen as inverters injecting power into the system, e.g. solar panels or wind turbines, and the remaining inverters act as consumers, which could, for example, be batteries that are being charged for later consumption.
One of the inverters serves as a slack node.
At this inverter, the phase is set to $\delta_{\mathrm{slack}}=0$, the voltage is kept at $E_{\mathrm{slack}}=1$, and the equations for the active and reactive power are not considered.
The slack node acts as an ideal voltage source that can provide an arbitrary amount of power.
Hence, we exclude the conditions in Eq.~\eqref{eq:stationary_state} for the slack node when calculating the fixed point.
If needed, the real and reactive power injections at the slack node can be computed afterwards.

As in the example for the single inverter discussed in Sec.~\ref{sec:single_inverter}, the desired frequency deviation is set to $\omega^d=0$. 
Since we are in a co-rotating reference frame, this corresponds to synchronized dynamics with the reference frequency.
The desired voltage was set to $E^d=1$, the recovery time to $\tau=0.1$, and the desired reactive power to $Q^d=0.05$.
In both cases, we will examine a scan over different values of desired power $P^d$, coupling strength $B$, and reactive droop gains $\chi$.

\begin{figure}
    \centering
    \includegraphics[width=\columnwidth]{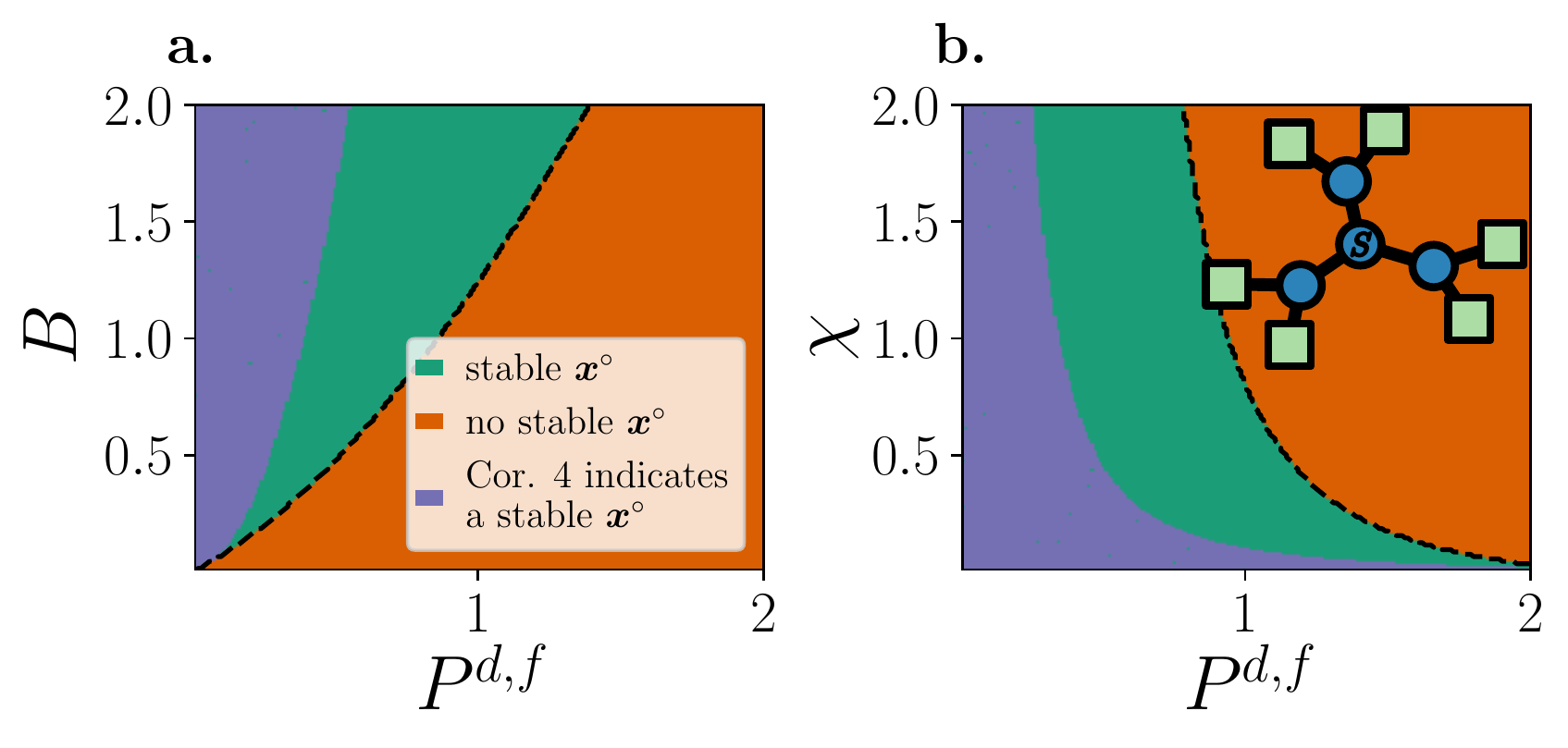}
    \caption{Corollary~\ref{cor:4}, a sufficient condition for stability, correctly identifies the region with stable fixed points for a considerable region in parameter space for a tree-like topology.
    The six outer nodes deliver power, while the inner four ones consume power.
    A slack node with fixed voltage magnitude $E_1=1$ and the reference power phase angle $\delta_1 = 0$ is located at the center (indicated by $S$).
    The reactive droop gain was chosen as $\chi = 0.5$ for the scan in the $B$-$P^{d,f}$ plane in panel \textbf{a.}, while the coupling strength $B=1.5$ was chosen for the scan in the $\chi$-$P^{d,f}$ plane in panel \textbf{b.}.
    Additionally, the coupling strength $\kappa=1$, desired voltage $E^d=1$, and desired reactive power $Q^d = 0.05$ were chosen. 
    The different colors show where the fixed point is stable or unstable, and where the corollaries indicate a stable fixed point. 
    Note, the region where Corollary~\ref{cor:4} indicates a stable fixed point always overlaps the region where the full Jacobian indicates a stable fixed point.}
    \label{fig:tree_inverter_stable_map}
\end{figure}

The two-inverter system has one inverter with a desired active power of $P_1^d=P^d$ and one with $P_2^d=-P^d$. 
Tracking the physical fixed points, determining their stability, and evaluating Corollary~\ref{cor:4} leads to the stability map that can be seen in Fig.~\ref{fig:two_inverter_stable_map}. 
For a given desired power $P^d$, a stable fixed point only exists if the coupling strength $B$ is sufficiently high and the reactive droop gain $\chi$ is kept relatively constrained.
The critical value of the coupling strength $B$ increases sub-linearly with the desired power $P^d$, while the critical reactive gain decreases with $P^d$.

We generally find a good agreement between the explicit stability criteria and the stability boundary resulting from the evaluation of the eigenvalues of the Jacobian.
In particular, Corollary~\ref{cor:4}, which represents a sufficient condition for stability, captures the general shape of the separatrix between stability and instability.
Notably, there is a small region where the fixed point is stable, while Corollary~\ref{cor:4} does not point to a stable fixed point.
This is to be expected as Corollary~\ref{cor:4} is merely a sufficient condition.

Subsequently, we investigate a system with $N=10$ nodes in a tree-like power grid (see inset in panel $\mathbf{b.}$ in Fig.~\ref{fig:tree_inverter_stable_map})
The outer nodes are net producers of power (e.g. solar panels) providing power to the four inner nodes that are net consumers of power.
While the outer nodes have a desired active power of $P^{d,f}$, the inner ones have a desired active power of $-3P^{d,f}/2$.
In this case, the central node acts as the slack node. 
The overall shape of the parameter region with a stable fixed point (see Fig.~\ref{fig:tree_inverter_stable_map}) is similar to the previously treated cases.
A minimum value of the coupling strength $B$ is required for a stable fixed point, where the critical value increases sub-linearly with $P^{d,f}$ (see panel $\mathbf{a.}$).
Furthermore, the reactive droop gains $\chi$ must not exceed a critical value, which decreases with increasing $P^{d,f}$ (see panel $\mathbf{b.}$).
In general, Corollary~\ref{cor:4} describes the shape of the stable parameter region well.
Note, in comparison with the two-inverter system (see Fig.~\ref{fig:two_inverter_stable_map}), the region where Corollary~\ref{cor:4} indicates a stable fixed point is smaller.
This highlights again that Corollary~\ref{cor:4} is a sufficient condition for a stable fixed point that conservatively predicts the stable parameter region.

In summary, the numerical simulations of the two considered test systems highlight the usefulness of the developed corollaries.
The developed corollaries highlight the role of different parameters, in particular, the reactive droop gains $\chi$, in shaping the parameter region with a stable fixed point.
As the corollaries were developed analytically, they show the interplay of these parameters in a general and transparent way.

\section{Conclusion and Discussion}
In this article, we investigated the collective dynamics of a network of droop-controlled inverters.
Modern power grids rely evermore on power-electronic devices given the increased penetration of renewable energy sources.
Since droop-controlled inverters are a promising type of power-electronic inverters, we need to understand how systems of connected droop-controlled inverters behave dynamically.
More specifically, we need to know the combination of values for the intrinsic parameters of each inverter that ensure the desired operating state or fixed point is stable.
We started by considering a single inverter coupled to an infinite grid. 
Using this simple setting, all fixed points and their stability could be determined analytically.
Scanning over the desired active power $P^d$, the reactive power $Q^d$, the coupling strength to the power grid $B$, and the reactive droop gain $\chi$ reveals when the single inverter can operate at a stable fixed point.  
The numerical results show that the existence of a stable fixed point necessitates a coupling strength $B$ that is sufficiently large to transmit the desired power $P^d$.
Furthermore, a stable fixed point can only be found if the chosen reactive droop gain $\chi$ does not exceed a critical value.

To understand the stability of droop-controlled inverters in extended networks, we examined the full set of equations of motion for arbitrary networks of inverters in a lossless setting. 
We decomposed the voltage and power-angle dynamics and employed linear stability analysis, resulting in the central stability conditions summarized in Lemma~\ref{Prop:Schur}.
Using these results, we were able to formulate several explicit stability criteria (Corollaries~\ref{cor:voltage1} to \ref{cor:5}).
Therein, a set of stability conditions are given as a set of inequalities, involving the isolated frequency and voltage subsystems as well as the full system.
Notably, connectivity bounds for the full system are tighter than for the isolated subsystems, as shown in Corollary~\ref{cor:voltage_stability_graph}. 
Furthermore, an upper bound for the reactive droop gains $\chi$ is given in Corollary~\ref{cor:4}. 
Since the droop gains can be programmed and chosen freely for each inverter, uncovering the upper bounds for these droop gains is needed to guarantee a stable operating state is obtained.

Subsequently, we numerically tested the derived stability criteria for two test systems by comparing the stability given by the full Jacobian to the one predicted by the derived inequalities.
In general, Corollary~\ref{cor:4} captures the shape of the parameter regions with a stable fixed point well.
The parameter region that is stable according to the full Jacobian is larger than the stable parameter region that is stable according to Corollary~\ref{cor:4}.
This discrepancy is larger for the larger test system network. 
Since Corollary~\ref{cor:4} is a sufficient condition for the stability of the considered fixed point, it is to be expected that it underestimates the size of the stable parameter region and therefore represents a conservative estimate.

Finally, the analytical treatment of droop-controlled inverters can be used to understand the dynamics of power grids that are driven by an increasing share of inverter-connected generation by solar and wind resources.
The insights obtained by the derived stability conditions can be used to test the parametric dependencies of the stability regions.
In particular, the analytic results highlight the interplay of the internal parameters of the inverters, such as the reactive droop gain $\chi$, and network properties, such as the algebraic connectivity $\lambda_2$.

\begin{acknowledgements}\vspace*{-1em}
We gratefully acknowledge support from the Helmholtz Association via the grant ``Uncertainty Quantification -- From Data to Reliable Knowledge (UQ)'' with grant no.~ZT-I-0029 and the Deutsche Forschungsgemeinschaft (DFG, German Research Foundation) via grant no.~ 491111487.
\end{acknowledgements}

\bibliography{bib}

\end{document}